%% file: main.tex
\documentclass[10pt,twocolumn,letterpaper]{article}

\usepackage[pagenumbers]{cvpr} 

\input{preamble}

\definecolor{cvprblue}{rgb}{0.21,0.49,0.74}
\usepackage{enumitem}
\usepackage{amsmath}
\usepackage{amssymb}
\usepackage{mathrsfs}
\usepackage{bm}
\usepackage{booktabs}
\usepackage{multirow}
\usepackage{threeparttable}
\usepackage[pagebackref,breaklinks,colorlinks,citecolor=cvprblue]{hyperref}

\def\paperID{3505} 
\def\confName{CVPR}
\def\confYear{2024}

\title{Building Bridges across Spatial and Temporal Resolutions: Reference-Based Super-Resolution via Change Priors and Conditional Diffusion Model}

\author{Runmin Dong$^{1,5}$, 
Shuai Yuan$^{2}$,
Bin Luo$^{1}$,
Mengxuan Chen$^{1,5}$,
Jinxiao Zhang$^{1,5}$,\\
Lixian Zhang$^{4,5*}$,
Weijia Li$^{3}$,
Juepeng Zheng$^{3,5}$,
Haohuan Fu$^{1,5}$\thanks{Corresponding authors}
\\
$^1$Tsinghua University~~~
$^2$The University of Hong Kong~~~
$^3$Sun Yat-Sen University ~~~\\
$^4$National Supercomputing Center in Shenzhen ~~~\\
$^5$Tsinghua University - Xi'an Institute
of Surveying and Mapping Joint Research Center~~~\\
{\tt\small drm@mail.tsinghua.edu.cn, haohuan@tsinghua.edu.cn}
}

\begin{document}
\maketitle
\input{sec/0_abstract}

\input{sec/1_introduction}

\input{sec/2_related_works}

\input{sec/3_methodology}

\input{sec/4_experiments}

\input{sec/5_conclusion}

\section*{Acknowledgements}

This research was supported in part by the National Natural Science Foundation of China (Grant No. T2125006 and No. 42301390), Jiangsu Innovation Capacity Building Program (Project No. BM2022028), China Postdoctoral Science Foundation (Grant No. 2023M731871), and Shuimu Tsinghua Scholar Project.

{
    \small
    \bibliographystyle{ieeenat_fullname}
    \bibliography{main}
}

\input{sec/X_suppl}

\end{document}

%% file: preamble.tex
\usepackage[dvipsnames]{xcolor}


%% file: sec/0_abstract.tex
\begin{abstract}
Reference-based super-resolution (RefSR) has the potential to build bridges across spatial and temporal resolutions of remote sensing images. However, existing RefSR methods are limited by the faithfulness of content reconstruction and the effectiveness of texture transfer in large scaling factors. Conditional diffusion models have opened up new opportunities for generating realistic high-resolution images, but effectively utilizing reference images within these models remains an area for further exploration. Furthermore, content fidelity is difficult to guarantee in areas without relevant reference information. To solve these issues, we propose a change-aware diffusion model named Ref-Diff for RefSR, using the land cover change priors to guide the denoising process explicitly. Specifically, we inject the priors into the denoising model to improve the utilization of reference information in unchanged areas and regulate the reconstruction of semantically relevant content in changed areas. With this powerful guidance, we decouple the semantics-guided denoising and reference texture-guided denoising processes to improve the model performance. Extensive experiments demonstrate the superior effectiveness and robustness of the proposed method compared with state-of-the-art RefSR methods in both quantitative and qualitative evaluations. The code and data are available at https://github.com/dongrunmin/RefDiff.
\end{abstract}

%% file: sec/1_introduction.tex
\begin{figure}[t]
\begin{center}
\vspace{-3mm}
\includegraphics[width=1.\linewidth]{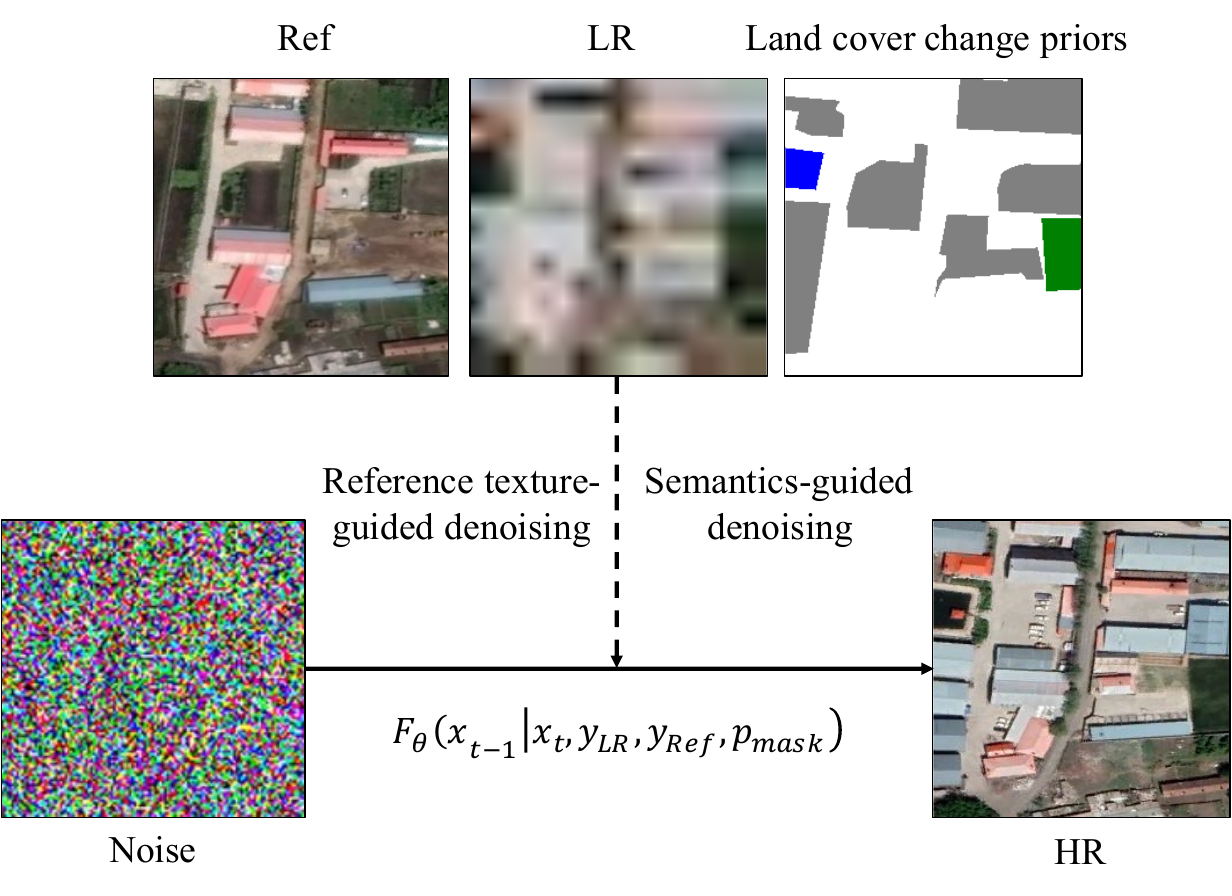}
\end{center}
   \caption{Illustration of the proposed change-aware diffusion model for RefSR. LR is a low-resolution image and HR is the corresponding high-resolution image. Ref represents a geographically matched reference high-resolution image acquired at another time.}
\label{fig:1}
\vspace{-4mm}
\end{figure}

\section{Introduction}
\label{sec:intro}

Spatiotemporal integrity of high-resolution remote sensing images is crucial for fine-grained urban management, long-time-series urban development study, disaster monitoring, and other remote sensing applications~\cite{jiang2020histif,dong2023large,zhang2024swcare}. However, due to limitations in remote sensing technologies and high hardware costs, we cannot simultaneously achieve high temporal resolution and high spatial resolution images on a large scale~\cite{luo2018stair,zhang2021making}. To tackle this issue, reference-based super-resolution (RefSR) can leverage geography-paired high-resolution reference (Ref) images and low-resolution (LR) images to integrate fine spatial content and high revisit frequency from different sensors~\cite{dong2023adaptive}. Although various RefSR methods achieve great progress, two major challenges remain to be solved for this scenario.

The first challenge is the land cover changes between Ref and LR images. Unlike the natural image domain, where Ref images are collected through image retrieval or captured from different viewpoints, Ref and LR images in remote sensing scenarios utilize geographic information to match the same location. Existing methods implicitly capture the land cover changes between LR and Ref images by adaptive learning or attention-based transformers~\cite{ma2022deep, RRSGAN}. However, the underuse or misuse problems of Ref information still exist in these methods. 

The second challenge is the large spatial resolution gaps between remote sensing sensors (e.g., 8$\times$ to 16$\times$). Existing RefSR methods are usually based on the generative adversarial network (GAN) and designed for a 4$\times$ scaling factor~\cite{ACMMM2023contrastive, zou2023geometry}. They can hardly reconstruct and transfer the details in the face of large-factor super-resolution. In recent years, conditional diffusion models have demonstrated greater effectiveness in image super-resolution and reconstruction than GAN~\cite{ho2022cascaded, xia2023diffir}. A straightforward way to boost RefSR is to use LR and Ref images as conditions for the diffusion model. To effectively utilize the reference information, some methods~\cite{HSRDIff-ICCV2023, li2023hyperspectral} inject Ref information into the blocks of the denoising networks. However, they implicitly model the relationship between LR and Ref images for denoising, leading to ambiguous usage of Ref information and content fidelity limitation.

To alleviate the above issues, we introduce land cover change priors to improve the effectiveness of reference feature usage and the faithfulness of content reconstruction (as shown in Figure~\ref{fig:1}). Benefiting from the development of remote sensing change detection (CD), we can use off-the-shelf CD methods to effectively capture land cover changes between images of different spatial resolutions~\cite{liu2022learning, zheng2021unsupervised, toker2022dynamicearthnet}. On the one hand, the land cover change priors enhance the utilization of reference information in unchanged areas. On the other hand, the changed land cover classes can guide the reconstruction of semantically relevant content in changed areas. Furthermore, according to the land cover change priors, we can decouple the semantics-guided denoising and reference texture-guided denoising in an iterative way to improve the model performance. To illustrate the effectiveness of the proposed method, we perform experiments on two datasets using two large scaling factors. Our method achieves state-of-the-art performance. In summary, our contributions are summarized as follows:

\begin{itemize}[label=\textbullet, leftmargin=2em] 
  \item We introduce the land cover change priors in RefSR to improve the content fidelity of reconstruction in changed areas and the effectiveness of texture transfer in unchanged areas, building bridges across spatial and temporal resolutions in remote sensing scenarios.
  \item We propose a novel RefSR method named Ref-Diff that injects the land cover change priors into the conditional diffusion model by the change-aware denoising model, enhancing the model's effectiveness in large-factor super-resolution. 
  \item Experimental results demonstrate that the proposed method outperforms the existing SOTA RefSR methods in both quantitative and qualitative aspects.

\end{itemize}

%% file: sec/2_related_works.tex
\section{Related Works}

\subsection{Reference-Based Super-Resolution Methods}  

Compared to single-image super-resolution (SISR), RefSR shows great potential in alleviating ill-posed problems and recovering realistic textures~\cite{Masa-sr-CVPR2021, DARTS2023}. Specifically, Jiang et al. ~\cite{C2-matching} propose a contrastive correspondence network and a teacher-student correlation distillation method to address the misalignment issues in the texture transfer and resolution gaps between LR and Ref images. RRSR~\cite{RRSR-ECCV2022} and AMSA~\cite{AMSA-AAAI2022} contribute to high-quality correspondence matching. Besides, Huang et al.~\cite{CVPR2022taskdecoupled} decouple super-resolution and texture transfer tasks to alleviate the issues of the underuse and misuse of Ref images. 

Owing to the pre-matching of LR and Ref images through geo-locations, existing RefSR methods for remote sensing images~\cite{RRSGAN, zhang2023reference} aim to transform relevant textures and suppress the irrelevant information fusion. However, their results contain apparent internal resolution inconsistencies between changed and unchanged regions in large-factor super-resolution. Because the details of changed regions can hardly be reconstructed using GAN-based methods. Therefore, recent works adopt the diffusion model to generate more realistic results~\cite{li2023hyperspectral}. For example, HSR-Diff~\cite{HSRDIff-ICCV2023} applies the conditional diffusion model and utilizes cross-attention as the conditioning mechanism to incorporate LR and Ref features into the denoising process, improving the perceptual quality. However, limited by implicit relationship modeling between LR and Ref images in the denoising process, the difficulty of denoising and the uncertainty of results are increased. In this work, we introduce the land cover change priors and explicitly use them to guide the denoising process.

\begin{figure*}[!t]
\vspace{-6mm}
\begin{center}
\includegraphics[width=1.03\linewidth]{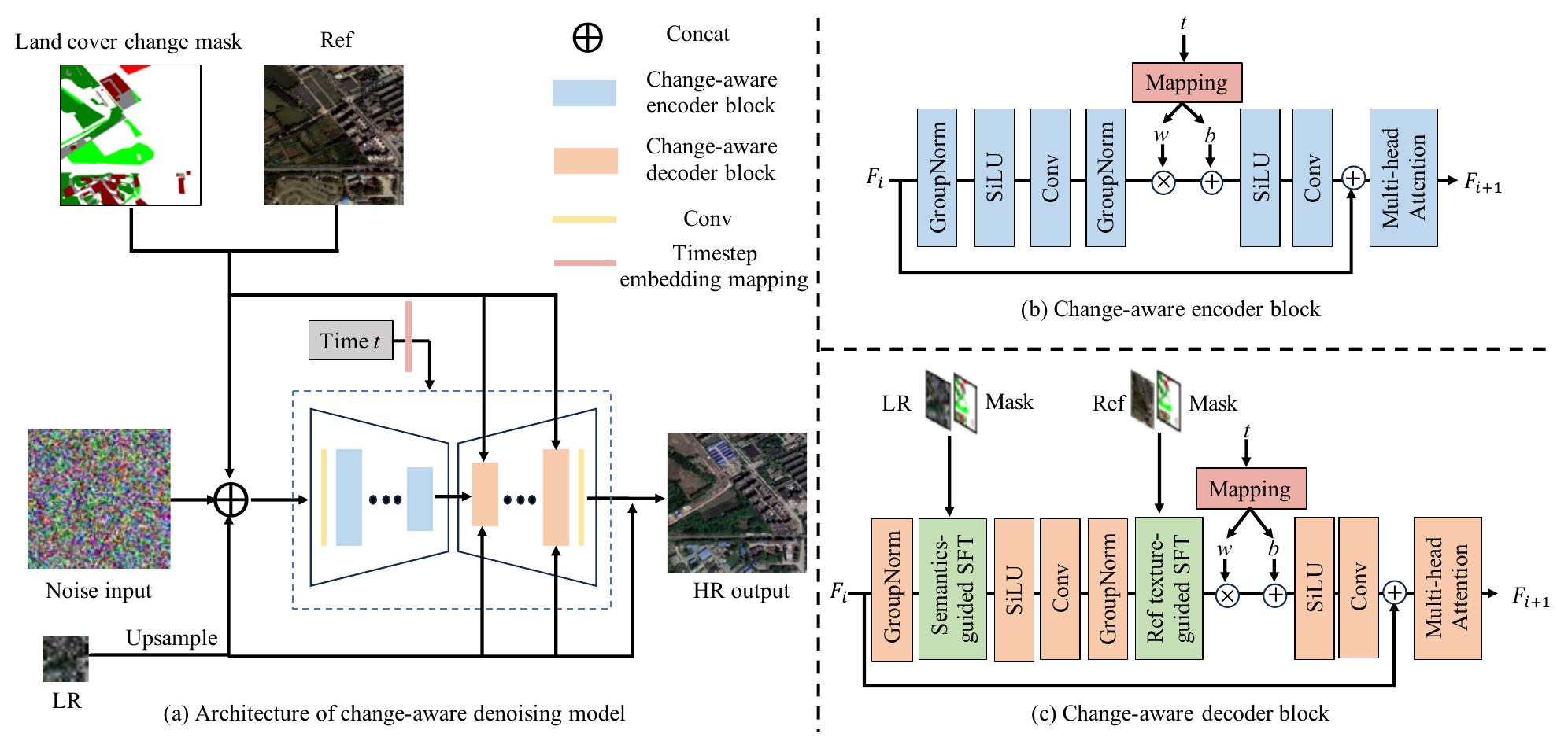}

\end{center}
   \caption{The architecture of the proposed change-aware denoising model. It consists of change-aware encoder and decoder blocks. The LR, Ref, and land cover change mask are combined with the noise input and are also injected into change-aware decoder blocks.
   }
\label{fig:2}
\vspace{-4mm}
\end{figure*}

\subsection{Conditional Diffusion Model for Super-Resolution} 

Benefiting from diffusion models, recent image super-resolution techniques have witnessed significant progress in terms of visual appeal and high-quality output. The initial works~\cite{SRDiff2022, SR32022-Google} utilize LR images as the condition for the diffusion processes to deal with the large-factor super-resolution. To further improve the effectiveness of image super-resolution, some works explore enhanced conditions to guide the denoising process. For example, ResDiff~\cite{shang2023resdiff} and ACDMSR~\cite{niu2023acdmsr} use the CNN-enhanced LR prediction as a condition to accelerate the generation process and acquire superior sample quality. BlindSRSNF~\cite{wu2022blind} and Dual-Diffusion~\cite{xu2023dual} combine degradation representations to the condition of the diffusion model to achieve satisfactory results in real-world scenarios. 

Except for simply combining those priors with the input of the conditional diffusion model, recent works integrate them into the denoising models~\cite{gao2023implicit, wang2023exploiting}. Wang et al.~\cite{wang2022semantic} introduce the semantic layout to the decoder by multi-layer spatially-adaptive normalization operators for semantic image synthesis. PASD~\cite{yang2023pixel} and DiffBIR~\cite{DiffBIR2023} adopt ControlNet to introduce priors like the high-level information extracted from CLIP or the enhanced LR representation by degradation removal. In this work, we explore the utilization of the land cover change priors in RefSR. We inject the priors into the denoising model, which decouples the semantics-guided and reference texture-guided denoising.

\subsection{Change Detection} 

With the development of change detection (CD) methods, existing works can achieve up to 80\% F1-score on different land cover categories~\cite{lv2022land}. For practical applications, recent CD models tend to be lightweight and can handle multi-temporal images with different resolutions~\cite{liu2022learning}. For example, Zheng et al.~\cite{zheng2021unsupervised} design a cross-resolution difference learning to bridge the resolution gap between two temporal images without resizing operations. Liu et al.~\cite{liu2021super} propose a SISR-based change detection network with a stacked attention module, achieving above 83\% F1-score in 8$\times$ resolution difference on the building CD task. These CD methods show high confidence and plug-and-play abilities, which can directly provide high-quality land cover change prior information for this work.

%% file: sec/3_methodology.tex
\section{Methodology}

In this paper, we adopt the conditional diffusion model to boost the effectiveness of RefSR methods in large scaling factors. To enhance the fidelity of the generated content and improve the effectiveness of the Ref image transfer, we introduce land cover change priors. The architecture of the proposed method is shown in Figure~\ref{fig:2}. We propose a novel change-aware denoising model, injecting the Ref features and the land cover change priors into the denoising blocks. Leveraging the priors, we decouple the semantics-guided denoising and reference texture-guided denoising processes in the decoder, and cope with the two processes iteratively.

\subsection{Preliminary}

The conditional diffusion model extends the basic diffusion model by incorporating conditions, including forward and reverse diffusion processes. Karras et al.~\cite{EDM2022} unify different diffusion models into the EDM framework. The training objective of EDM is defined as:

\begin{equation}
\mathbb{E}_{{\sigma},\bm{\textbf{y}},\bm{\textbf{n}}}[\lambda(\sigma)||D(\bm{\textbf{y}}+\bm{\textbf{n}};\sigma)-\bm{\textbf{y}}||_2^2],
\label{eq:1}
\end{equation}

\noindent where standard deviation $\sigma$ controls the noise level, $\textbf{y}$ is a training image and $\textbf{n}$ is noise. $D(\cdot)$ is a denoiser function. $\lambda(\sigma)$ is the loss weight. 

To effectively train a neural network, the preconditioning of EDM is defined as:

\begin{equation}
D_{\theta}(\bm{\textbf{x}};\sigma)=c_{\textrm{skip}}(\sigma)\bm{\textbf{x}}+c_{\textrm{out}}(\sigma)F_{\theta}(c_{\textrm{in}}(\sigma)\bm{\textbf{x}};c_{\textrm{noise}}(\sigma)),
\label{eq:2}
\end{equation}

\noindent where $\textbf{x}=\textbf{y}+\textbf{n}$. $F_{\theta}(\cdot)$ represents the neural network undergoing training. $c_{\textrm{skip}}$ adjusts the skip connection. $c_{\textrm{in}}$ and $c_{\textrm{out}}$ scale the input and output magnitudes, respectively. $c_{\textrm{noise}}$ is used to map the noise level $\sigma$ into a conditioning input for $F_{\theta}(\cdot)$. In this work, the diffusion architecture follows the formulations of the training objective, preconditioning, and other implementations in EDM.

\subsection{Change-Aware Denoising Model}

This work aims to exploit the land cover change priors to facilitate RefSR in large scaling factors for remote sensing images. The proposed change-aware denoising model is shown in Figure~\ref{fig:2}(a). Inspired by~\cite{park2019semantic,wang2022semantic}, the land cover change prior can be regarded as the semantic layout of the changed areas between LR and Ref images for the semantics-guided denoising. Meanwhile, the texture details can be enhanced in the unchanged areas through reference texture-guided denoising. As a result, the semantics-guided denoising and reference texture-guided denoising processes can be decoupled in the change-aware decoder (see Figure~\ref{fig:2}(c)), further improving the denoising results.

\noindent\textbf{Land Cover Change Priors.} Land cover change priors used in this work are the pixel-level multi-category change detection mask for each image pair, including a no-change class and different land cover change classes. To fully unleash the potential of land cover priors in training, we use the ground truth of the land cover change mask as the condition. In real applications, change detection masks can be generated by the off-the-shelf end-to-end change detection methods or two-stage land cover classification methods. As shown in Figure~\ref{fig:2}(a), the land cover change mask is combined with the noise as input and also injected into the change-aware decoder.

\noindent\textbf{Change-Aware Encoder.} To improve the computational performance and avoid over-intervention of LR denoising, the LR image, Ref image, and land cover change mask are concatenated with the noisy image as the input to the encoder, instead of being injected into the encoder blocks as in \cite{HSRDIff-ICCV2023}. The architecture of the encoder is based on the improved U-Net in~\cite{EDM2022} (see Figure~\ref{fig:2}(b)). Each change-aware encoder block consists of group normalization, convolution, SiLU, and a multi-head attention module. Since each timestep $t$ corresponds to a certain noise level, we map the timestep embedding into learnable weight $w(t)$ and bias $b(t)$ to regulate the features. Multi-head attention runs through the attention process multiple times in parallel, each with its own set of learnable parameters~\cite{vaswani2017attention}.

\noindent\textbf{Change-Aware Decoder.} As shown in Figure~\ref{fig:2}(c), we inject the features of land cover change masks and Ref images into the change-aware decoder blocks. With the land cover change priors, we decouple the semantics-guided denoising in the changed areas and reference texture-guided denoising in the unchanged areas. To tackle the mislabel problem in the land cover change priors, we combine the land cover change masks and LR images for the semantics-guided spatial feature transform (SFT) module. Considering the no-change class in land cover change mask can guide the utilization of Ref texture, we combine the land cover change masks and Ref images for the Ref texture-guided SFT module. In this way, the denoising of changed and unchanged areas can reinforce each other by an iterative solution. Considering the accuracy of predicting land cover change masks through change detection methods is usually between 60\% to 80\% in practical applications, we combine the guidance features and denoising features for the learning of spatially adaptive weight and bias, rather than only use the guidance features like the original SFT~\cite{SFT2018} and SPADE~\cite{park2019semantic,wang2022semantic}. The modified SFT module can be formulated as:

\begin{equation}
F_{i+1}=\gamma_{i}(F_{e} \oplus F_{i}) \cdot F_{i} + \beta_{i}(F_{e} \oplus F_{{i}}), 
\label{eq:3}
\end{equation}

\noindent where $F_{i}$ and $F_{i+1}$ are the input and output features of the SFT module, respectively. $\gamma_{i}(\cdot)$, $\beta_{i}(\cdot)$ are the spatially-adaptive weight and bias learned from the combination of the guidance features $F_{e}$ obtained by the extractor and the input features $F_{i}$, respectively.

\subsection{Degradation Model and Implementation Details}

We adopt a comprehensive degradation to simulate LR images in real-world scenarios for training. According to off-the-shelf blind super-resolution methods~\cite{gu2019blind, wang2021real} and the characteristics of remote sensing sensors~\cite{dong2022real, qiu2023cross}, we adopt isotropic Gaussian blur, anisotropic Gaussian blur, motion blur, resize with different interpolation methods, additive Gaussian noise, and JPEG compression noise to synthesis LR images. The setting of degradation complexity is based on the scaling factor. In the experiments, the degradation model for $16\times$ datasets is simpler than that for $8\times$ datasets.

During training, each high-resolution (HR) image, Ref image, and land cover change mask are randomly cropped to a size of $256\times256$, and the size of the corresponding LR image is associated with the scaling factors. The implementation of the diffusion model is according to~\cite{EDM2022}. We utilize a dropout rate of 0.2. The batch size is set to 48. We use the Adam optimizer with $\beta_{1} = 0.9$ and $\beta_{2} = 0.999$. The learning rate is initialized as $1\times{10}^{-4}$. The model is updated for 500k iterations using 4 NVIDIA A800 GPUs.

%% file: sec/4_experiments.tex
\begin{figure*}[t]
\begin{center}
\includegraphics[width=1.\linewidth]{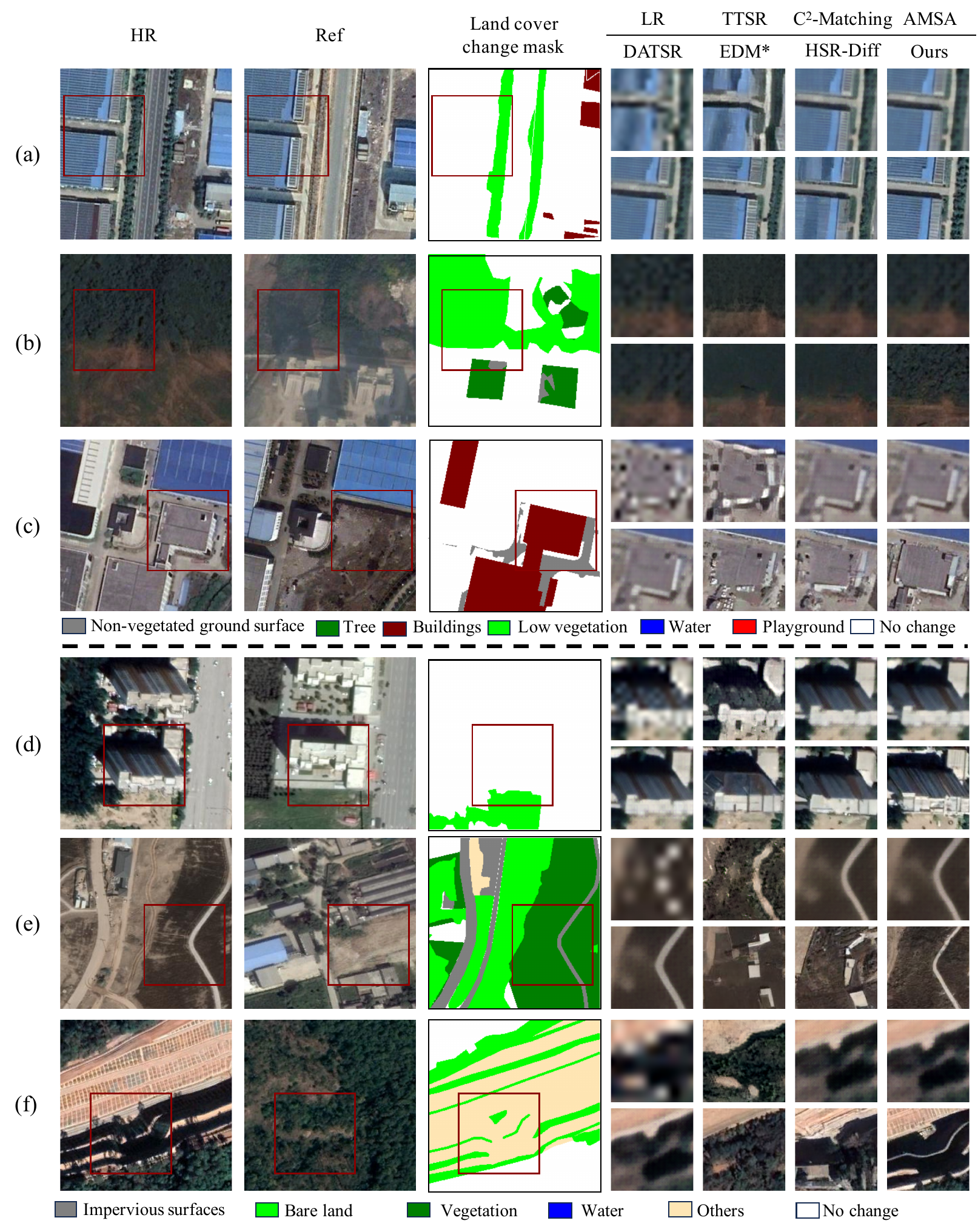}
\end{center}
\vspace{-6mm}
\caption{Comparison results on SECOND (a-c) and CNAM-CD (d-f) datasets with $8\times$ and $16\times$ scaling factors.}
\label{fig:3}
\vspace{-6mm}
\end{figure*}

\begin{table*}[t]
\caption{Quantitative comparison with different reference-based methods on two datasets, i.e., SECOND and CNAM-CD. Each dataset is evaluated at two large scaling factors, i.e., $8\times$ and $16\times$. Lower LPIPS and FID values indicate better results. Bold indicates the best results.}
\begin{center}
\vspace{-2mm}
\begin{threeparttable}
\begin{tabular}{c|cc|cc|cc|cc}
\hline
\multirow{3}[1]{*}{Methods} & \multicolumn{4}{c|}{SECOND} & \multicolumn{4}{c}{CNAM-CD} \\
\cline{2-9}    \multicolumn{1}{c|}{} & \multicolumn{2}{c|}{8$\times$} & \multicolumn{2}{c|}{16$\times$} & \multicolumn{2}{c|}{8$\times$} & \multicolumn{2}{c}{16$\times$} \\
\cline{2-9}    \multicolumn{1}{c|}{} & \multicolumn{1}{c}{LPIPS$\downarrow$} & \multicolumn{1}{c|}{FID$\downarrow$} & \multicolumn{1}{c}{LPIPS$\downarrow$} & \multicolumn{1}{c|}{FID$\downarrow$} & \multicolumn{1}{c}{LPIPS$\downarrow$} & \multicolumn{1}{c|}{FID$\downarrow$} & \multicolumn{1}{c}{LPIPS$\downarrow$} & \multicolumn{1}{c}{FID$\downarrow$} \\
    \hline
    TTSR~\cite{TTSR} & 0.3799 & 142.4030 & 0.4743 & 232.0805 & 0.4041 & 163.8194 & 0.4700  & 191.8991 \\
    WTRN~\cite{WTRN-TIP2022} & 0.5081 & 110.4582 & 0.8426 & 260.4063 & 0.4915 & 142.1856 & 0.8121 & 283.1182 \\
    $\textrm{C}^2$-Matching~\cite{C2-matching} & 0.3351 & 62.2991 & 0.4972 & 123.3611 & 0.3697 & 97.9389 & 0.5494 & 182.0406 \\
    AMSA~\cite{AMSA-AAAI2022} & 0.3601 & 56.9689 & 0.5353 & 135.6814 & 0.3989 & 92.3733 & 0.5613 & 169.0448 \\
    DATSR~\cite{DATSR-ECCV2022} & 0.3525 & 56.0531 & 0.5078 & 111.7564 & 0.3894 & 94.7989 & 0.5331 & 163.8926 \\
    \hline
    EDM$^{*}$~\cite{EDM2022} & 0.2886 & 34.5802 & 0.3440 & 37.5573 & 0.3301 & 53.1511 & 0.3902 & 59.4250 \\
    HSR-Diff~\cite{HSRDIff-ICCV2023} & 0.2689 & 45.4743 & 0.3437 & 51.1473 & 0.3045 & 67.2524 & 0.3771 & 68.1954 \\
    Ref-Diff (ours) & \textbf{0.2642} & \textbf{32.5961} & \textbf{0.3433}  & \textbf{33.9690} & \textbf{0.2791} & \textbf{43.0152} & \textbf{0.3519} & \textbf{45.6511} \\
    \hline
\end{tabular}
\begin{tablenotes}
        \footnotesize
        \item[1] EDM$^{*}$ represents that the original method is modified for the RefSR task, which combines LR and Ref images with the noise input.
        \end{tablenotes}
\end{threeparttable}
\end{center}
\label{tab:1}
\vspace{-4mm}
\end{table*}

\section{Experiments}

\subsection{Datasets and Evaluation}

\noindent\textbf{SECOND Dataset.} SECOND~\cite{SECOND} is a semantic change detection dataset with 7 land cover class annotations, including non-vegetated ground surface, tree, low vegetation, water, buildings, playgrounds, and unchanged areas. The images are collected from different sensors and areas with resolutions between 0.5 and 1 meters, guaranteeing style diversity and scene diversity. In this work, we use 2,668 image pairs with a size of $512 \times 512$ for training and 1,200 image pairs with a size of $256\times256$ for testing. 

\noindent\textbf{CNAM-CD Dataset.} CNAM-CD~\cite{CNAM-CD} is a multi-class change detection dataset with a resolution of 0.5 meter, including 6 land cover classes, i.e., bare land, vegetation, water, impervious surfaces (buildings, roads, parking lots, squares, etc.), others (clouds, hard shadows, clutter, etc.), and unchanged areas. The image pairs are collected from Google Earth from 2013 to 2022. We use 2,258 image pairs with a size of $512 \times 512$ for training and 1,000 image pairs with a size of $256\times256$ for testing. 

\noindent\textbf{Evaluation.} The performance of the proposed approach and other competing methods on test datasets are assessed with the learned perceptual image patch similarity (LPIPS) and Fréchet inception distance (FID), which can better quantify both fidelity and perceptual quality than PSNR and SSIM~\cite{yang2023pixel}. LPIPS~\cite{LPIPS} is a full-reference metric designed to capture the perceptual quality of images. It measures the similarity between two images based on their perceptual features obtained by a pre-trained deep network which is the AlexNet model~\cite{alexnet} in the experiments. FID~\cite{FID} quantifies the similarity between the distributions of features extracted from a pre-trained Inception network for real and generated images. Lower LPIPS and FID values imply better results.

\begin{table*}[t]
\caption{Ablation study of our method on SECOND $8\times$ dataset. Lower LPIPS and FID values indicate better results.}
\vspace{-2mm}
\begin{center}
\begin{tabular}{ccccccc}
 \hline
\multicolumn{1}{c}{LR } & \multicolumn{1}{c}{Ref } & \multicolumn{1}{c}{Land cover change mask} & \multicolumn{1}{c}{Ref texture-guided } & \multicolumn{1}{c}{Semantics-guided} & \multicolumn{1}{c}{\multirow{2}[2]{*}{LPIPS$\downarrow$}} & \multicolumn{1}{c}{\multirow{2}[2]{*}{FID$\downarrow$}} \\
\multicolumn{1}{c}{condition} & \multicolumn{1}{c}{condition} & \multicolumn{1}{c}{condition} & \multicolumn{1}{c}{SFT} & \multicolumn{1}{c}{ SFT} &       &  \\
\hline
          \checkmark &       &       &       &       & 0.3115 & 41.8340 \\
          \checkmark &       \checkmark &       &       &       & 0.2886 & 34.5802 \\
          \checkmark &       \checkmark &       \checkmark &       &       & 0.2785 & 34.0638 \\
          \checkmark &       \checkmark &       \checkmark &       \checkmark &       & 0.2709 & 33.7583 \\
          \checkmark &       \checkmark &       \checkmark &       &       \checkmark & 0.2723 & 33.6805 \\
          \checkmark &       \checkmark &       \checkmark &       \checkmark &       \checkmark & \textbf{0.2642} & \textbf{32.5961} \\
\hline
\end{tabular}
\end{center}
\label{tab:2}
\vspace{-4mm}
\end{table*}

\begin{table*}[t]
\caption{Results using land cover change predictions.}
\vspace{-2mm}
\begin{center}
\begin{tabular}{c|c|ccc|cc}
\hline 
Dataset                       & Land cover change mask & F1$\uparrow$    & Precision$\uparrow$ & Recall$\uparrow$ & LPIPS$\downarrow$  & FID$\downarrow$     \\ \hline 
\multirow{2}{*}{SECOND 8X}    & GT                     & -     & -         & -      & 0.2642 & 32.5961 \\
                              & Prediction             & 87.72 & 86.41     & 86.30  & 0.2657 &  33.1453       \\ 
\multirow{2}{*}{SECOND 16X}   & GT                     & -     & -         & -      & 0.3433 & 33.9690 \\
                              & Prediction             & 84.94 & 84.23     & 84.70  & 0.3404 & 35.0477 \\ \hline
\multirow{2}{*}{CNAM-CD  8X}  & GT                     & -     & -         & -      & 0.2791 & 43.0152 \\
                              & Prediction             & 87.11 & 87.47     & 85.81  &    0.3159    &    48.3315     \\ 
\multirow{2}{*}{CNAM-CD  16X} & GT                     & -     & -         & -      & 0.3519 & 45.6511 \\
                              & Prediction             & 87.20 & 84.60     & 85.01  &    0.3889    &  56.7857   \\ \hline     
\end{tabular}
\end{center}
\label{tab:3}
\vspace{-4mm}
\end{table*}

\subsection{Comparison Results}

The proposed method is compared with existing GAN-based and diffusion model-based RefSR methods on two datasets with two large scaling factors (i.e., $8\times$ and $16\times$). The compared RefSR methods include five GAN-based methods (i.e., TTSR~\cite{TTSR}, WTRN~\cite{WTRN-TIP2022}, $C^{2}$-Matching~\cite{C2-matching}, AMSA~\cite{AMSA-AAAI2022}, and DATSR~\cite{DATSR-ECCV2022}), and two diffusion model-based RefSR methods (i.e., EDM$^{*}$~\cite{EDM2022} and HSR-Diff~\cite{HSRDIff-ICCV2023}). EDM$^{*}$ represents that the original method is modified for the RefSR task, which combines LR and Ref images with the noise input. For a fair comparison, the LR images are synthesized by bicubic interpolation. 

Table~\ref{tab:1} shows the quantitative comparison results. Our method achieves the best LPIPS and FID performance in four sets of comparison experiments, demonstrating the advanced fidelity and perceptual quality of our results. According to the comparison results between GAN-based and diffusion model-based RefSR methods, the latter shows more powerful ability to bridge the resolution gap in large scaling factors. Owing to the utilization of the land cover changes priors, the proposed method performs consistently better than the other two diffusion model-based RefSR methods.

We further present the visual comparison in Figure~\ref{fig:3}. Figure~\ref{fig:3}(a) shows an example with slight changes on the SECOND $8\times$ dataset. Our method can effectively transfer the relevant textures from the Ref image to the LR image in the unchanged areas, while the competing methods suffer from artifacts or blurred results. In the meantime, the vegetation reconstruction results are more realistic than comparison results in the changed areas. Figure~\ref{fig:3}(b) further demonstrates the effectiveness of vegetation reconstruction. Figure~\ref{fig:3}(c) shows an example with dramatic changes on the SECOND $16\times$ dataset. The HR image contains a building where the non-vegetated ground surface is in the Ref image. The proposed method can guarantee the faithfulness of content reconstruction, while other methods cannot cope with this challenging scenario. 

Similarly, Figure~\ref{fig:3}(d) shows an example with slight changes on the CNAM-CD $8\times$ dataset. The texture of the building in our results is more realistic than other results, demonstrating the effectiveness of feature alignment and texture transfer in our method. Figure~\ref{fig:3}(e) shows an example with dramatic changes on the CNAM-CD $16\times$ dataset. Our method correctly reconstructs the road and vegetation with the guidance of land cover change priors. The other two diffusion model-based RefSR methods produce illusory buildings and confusing layouts due to limited priors. Although remaining content fidelity, the GAN-based RefSR methods cannot reconstruct texture details, resulting in the spatial resolution gap between super-resolution results and HR images. Figure~\ref{fig:3}(f) also exhibits a dramatic change area on the CNAM-CD $16\times$ dataset. The results of our method remain faithful to content reconstruction and can even remove the tree shadow in LR images. 

In summary, the proposed method improves the content fidelity of reconstruction in changed areas and the effectiveness of texture transfer in unchanged areas, effectively building bridges across spatial and temporal resolutions.

\subsection{Ablation Study}

We perform ablation study on the SECOND 8$\times$ dataset to verify the effectiveness of the proposed method. In turn, we add the LR image, Ref image, and land cover change mask to the input of the conditional diffusion model. As shown in Table~\ref{tab:2}, using the Ref condition can largely improve the diffusion model ability in the super-resolution task, which is a promising way to narrow the gap between spatial resolutions in remote sensing scenarios. Taking the land cover change mask as a condition can further enhance the results. Still, the improvements are limited due to the simple combination between the conditions and the noise input. 

We further conduct three experiments to verify the effectiveness of the semantics-guided SFT module, the Ref texture-guided SFT module, and the decoupled denoising strategy which uses both SFT modules. The results in Table~\ref{tab:2} show that using either SFT module improves the denoising results because the guidance information further modulates the features in the decoder. Besides, building upon the two types of SFT modules, the decoupled denoising strategy with the iterative semantics-guided denoising and reference texture-guided denoising obtains the best results. Besides, we provide the ablation study of enhanced spatial feature transform module and results of real scenarios in the supplementary.

\begin{figure*}[t]
\begin{center}
\vspace{-4mm}
\includegraphics[width=1.\linewidth]{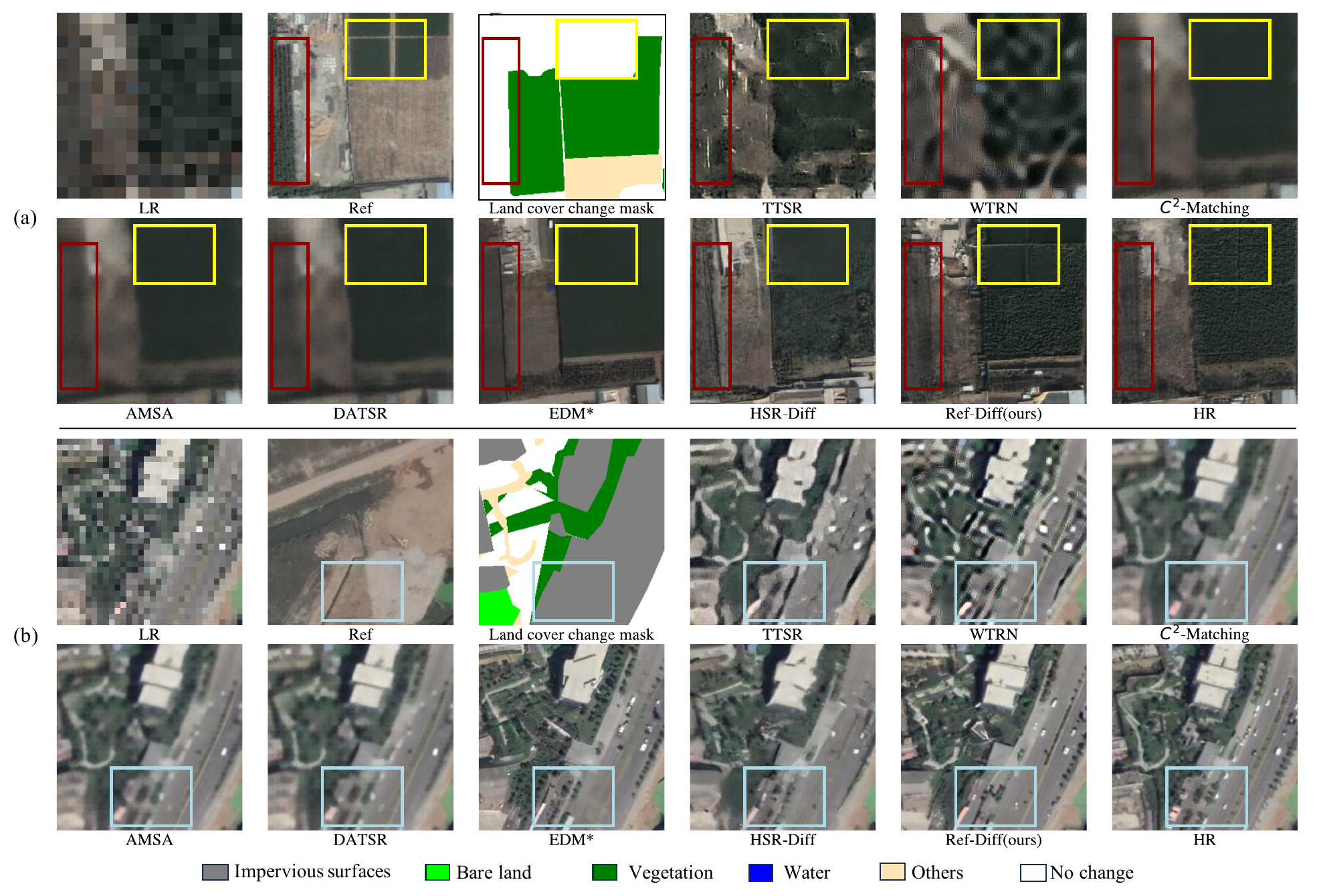}
\end{center}
   \caption{The results for two examples with mislabeled land cover change masks on CNAM-CD 8$\times$ and 16$\times$ datasets. (a) shows an example with false negative detection, and (b) shows an example with false positive detection.}
\label{fig:4}
\vspace{-2mm}
\end{figure*}

\subsection{Experiments Using Land Cover Change Predictions} 

We conduct experiments to illustrate the impact of utilizing predicted land cover change masks. Specifically, we train change detection (CD) models based on ChangeFormer~\cite{bandara2022transformer}. Combining a structured transformer encoder and an MLP decode, ChangeFormer is an ideal plug-and-play CD network for this work. Table~\ref{tab:3} presents the quantitative comparison results using the prediction of land cover change masks. Although the performance is reduced compared to using ground truth (GT), our method still outperforms the comparison methods (refer to Table~\ref{tab:1}). This reinforces the effectiveness of using the proposed method in real scenarios.

\subsection{Discussion of the Interaction between RefSR and CD Tasks} 

Ideally, accurate land cover change priors improve the confidence of Ref texture transfer in unchanged areas and content generation in changed areas. However, in practical scenarios, the utilization of land cover change priors may introduce misleading information due to mislabeling issues or prediction errors in CD tasks. Figure~\ref{fig:4} illustrates two common CD errors, i.e., false negatives (FN) and false positives (FP). The FN issue may lead to the introduction of false textures into the results, as depicted in Figure~\ref{fig:4}(a). Figure~\ref{fig:4}(b) presents an example of the FP issue, where vegetated areas are incorrectly labeled as impervious surfaces. The FP issue also undermines our results. Therefore, the improvement of CD accuracy will enhance the change-aware RefSR results. Moreover, a fine-grained classification system of land cover change will further facilitate the fidelity of content reconstruction in RefSR. 

On the other hand, this work demonstrates the potential of change-aware RefSR in synthesizing well-labeled change detection data by semantic layout control and LR image collection. Consequently, RefSR and CD tasks can mutually reinforcement each other.

%% file: sec/5_conclusion.tex
\section{Conclusion}

In this work, we propose a change-aware diffusion model for reference-based remote sensing image super-resolution to improve the faithfulness of content reconstruction and the effectiveness of texture transfer in large scaling factors. We inject the land cover change priors into the conditional diffusion model to explicitly guide denoising. With this powerful guidance, we decouple the semantic-guided denoising process in changed areas and the reference texture-guided denoising process in unchanged areas. We achieve the best quantitative and qualitative over state of the arts. This work also demonstrates the potential for mutual reinforcement between RefSR and change detection tasks. In future work, we will integrate change detection methods into the RefSR framework to enhance practicability.

%% file: sec/X_suppl.tex
\clearpage
\setcounter{section}{0}
\maketitlesupplementary
\renewcommand\thesection{\Alph {section}}
\renewcommand\thefigure{\Alph{section}\arabic{figure}} 
\renewcommand\thetable{\Alph{section}\arabic{table}}

\section{Ablation Study of Enhanced Spatial Feature Transform Module}
Existing approaches, such as the spatial feature transform (SFT)~\cite{SFT2018} and SPADE~\cite{park2019semantic,wang2022semantic}, leverage an ideal semantic map to regulate denoising features through its embedding. To enhance the robustness of utilizing land cover change priors, we introduce an enhanced SFT module for semantic guidance and reference (Ref) texture guidance. The enhanced SFT module integrates guidance features with denoising features to regulate the latter. Additionally, we incorporate the low-resolution (LR) image with the land cover change mask for semantics-guided SFT, going beyond the use of only semantic embedding. The results in Table~\ref{tab:4} illustrate the effectiveness of the enhanced SFT module in alleviating the negative impact of imprecise priors.

\section{Experiments in Real Scenarios}

To illustrate the effectiveness of the proposed method, we evaluate its performance in real scenarios using two distinct datasets located in areas different from the training datasets. The first dataset is situated in Jiaxing, China, employing a classification system identical to the SECOND dataset. The second dataset is derived from the HRSCD dataset~\cite{daudt2019multitask}, covering two regions in France (i.e., Rennes and Caen). The classification system of this dataset can be roughly mapped to that in the CNAM-CD dataset.Real LR and high-resolution (HR) images are collected from Google Earth Engine based on the geographical information of reference images. Real LR and HR images include Google Earth images with different levels of RGB bands. The resolution of real HR and Ref images is 0.5 meter. Note that the real LR images and real HR images may be captured by different sensors. 

\begin{figure*}[t]
\begin{center}
\includegraphics[width=0.83\linewidth]{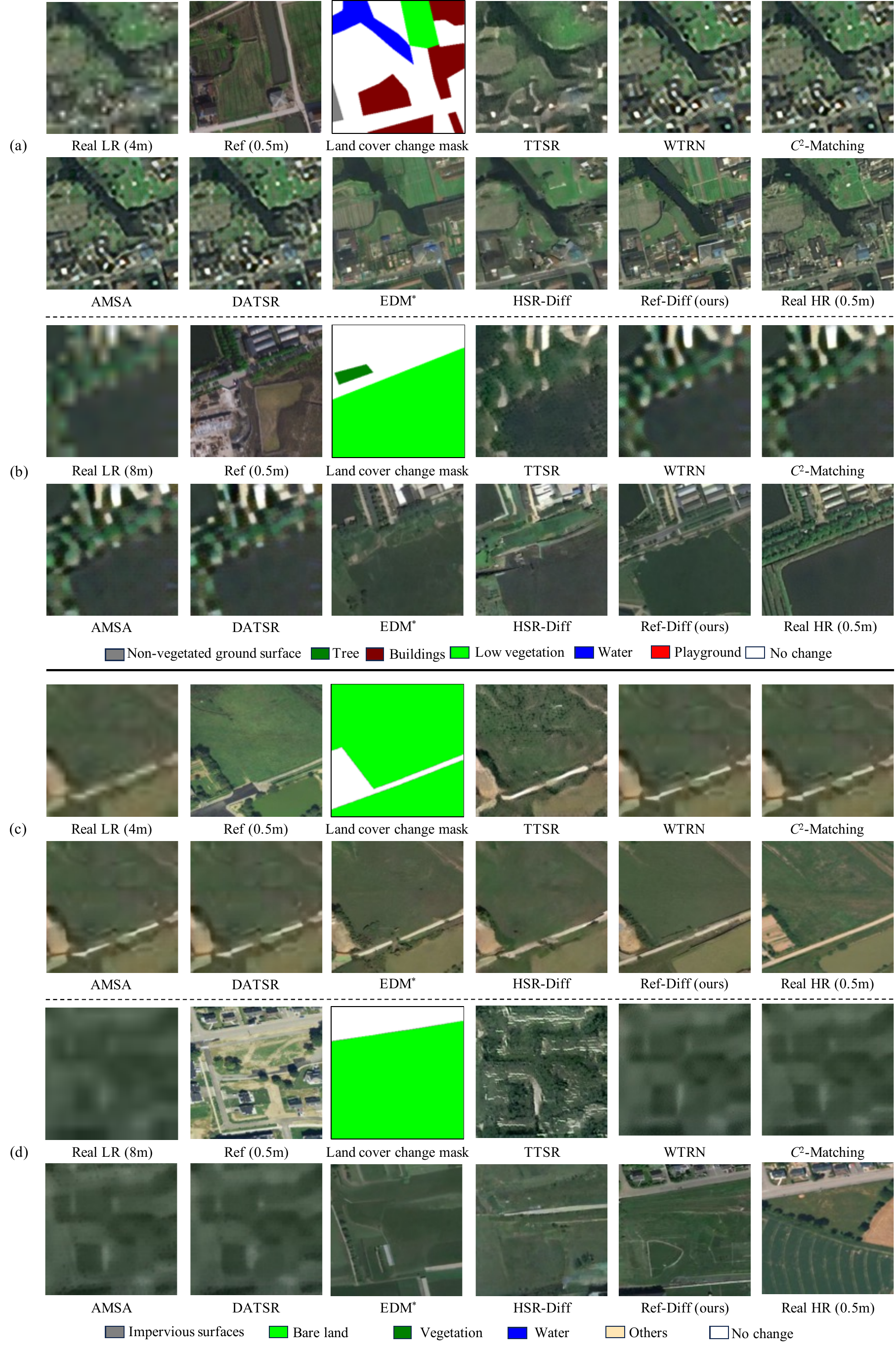}
\end{center}
\vspace{-6mm}
\caption{Comparison results on two real datasets. (a-b) are located in Jiaxing, China. (c-d) are located in Rennes and Caen, France. (a) and (c) are with $8\times$ scaling factor. (b) and (d) are with $16\times$ scaling factor.}
\vspace{-6mm}
\label{fig:5}
\end{figure*}

As depicted in Figure~\ref{fig:5}, our method outperforms competing RefSR methods, showcasing superior visual results on the two real datasets. It demonstrates the advanced fidelity and perceptual quality achieved by our method.

\section{Further Analysis of the Proposed Method} 

We summarize the shortcomings of our method in three aspects. Firstly, it is challenging for the proposed method to reconstruct small objects such as vehicles. Secondly, our method cannot handle the super-resolution in substantial scaling factors (e.g., $32\times$). Finally, diffusion model-based RefSR is time-consuming compared with GAN-based RefSR methods. But this issue is promising to be optimized by recent studies on diffusion model acceleration such as~\cite{xia2023diffir}.

In our future work, we will further explore the end-to-end method, integrating the change detection methods into RefSR models, to improve the practicality of our method. As analyzed in Section 4.4 in the main text, the simple two-stage strategy of cascading a change detector and our RefSR model does not fully unleash the capabilities of the proposed method in real applications. An end-to-end approach is expected to bridge this gap. Additionally, we will also apply different treatments to different categories of land cover change. Our method is designed to adaptively handle various types of changes. Given that the difficulty of reconstruction varies among land cover change types, introducing an explicit constraint or guidance could prove beneficial in further improving the results.

\begin{table}[t]
\caption{Results using the original SFT and enhanced SFT modules. Bold indicates the best results.}
\begin{center}
\begin{tabular}{c|cc}
\hline 
Method     & LPIPS$\downarrow$    & FID$\downarrow$   \\ \hline 
With the original SFT & 0.2725          & 33.9478      \\
With the enhanced SFT  & \textbf{0.2642} & \textbf{32.5961}  \\ \hline 
\end{tabular}
\end{center}
\label{tab:4}
\end{table}